\documentclass[preprint,aps,newabstract]{revtex4}
\usepackage{graphicx}
\newcommand{\lsim}{\mathrel{\mathop{\kern 0pt \rlap
  {\raise.2ex\hbox{$<$}}}
  \lower.9ex\hbox{\kern-.190em $\sim$}}}
\newcommand{\gsim}{\mathrel{\mathop{\kern 0pt \rlap
  {\raise.2ex\hbox{$>$}}}
  \lower.9ex\hbox{\kern-.190em $\sim$}}}
\newcommand{\beq}     {\begin{equation}}
\newcommand{\eeq}     {\end{equation}}
\newcommand{\bea}     {\begin{eqnarray}}
\newcommand{\eea}     {\end{eqnarray}}
\newcommand{\intfive}      {\int d^5 x \sqrt{-G}}
\newcommand{\intfour}      {\int d^4 x }
\newcommand{\dt}      {\delta}
\newcommand{\al}      {\alpha}
\newcommand{\be}      {\beta}
\newcommand{\C}      {C^{btW}}
\newcommand{\Cu}      {C^{bu W}}
\newcommand{\n}      {{(n)}}
\newcommand{\m}      {{(m)}}
\newcommand{\es}      {\epsilon}
\newcommand{\ev}      {\equiv}

\newcommand{\lm}      {\lambda}

\newcommand{\gm}      {\gamma}
\newcommand{\sg}      {\sigma}
\newcommand{\kp}      {\kappa}

\newcommand{\Lg}       {{\mathcal L}}
\newcommand{\M}       {{\mathcal M}}
\newcommand{\D}       {{\mathcal D}}
\newcommand{\no}      {\nonumber}
\begin{document}
\draft

\preprint{KIAS P02017}

\title{
Top quark Kaluza-Klein mode mixing \\
in the Randall-Sundrum bulk Standard Model  \\
and $B\to X_s \gamma$
}

\author{
C.~S. Kim, J.~D. Kim}
\affiliation{Department of Physics and
IPAP,
 Yonsei University, Seoul 120-749, Korea}
\author{Jeonghyeon Song
}
\affiliation{School of Physics,
Korea Institute for Advanced Study\\
207-43 Cheongryangri-dong,
Dongdaemun-gu, Seoul 130-012, Korea}

\begin{abstract}
We study top quark Kaluza-Klein (KK) mode mixing in the
Randall-Sundrum  scenario with all the SM fermions and gauge
bosons in the bulk. Even though the simple assumption of universal
bulk fermion mass $m_\psi$ leads to the same KK mass spectrum for
all the SM fermions and thus suppresses new contributions to the
flavor changing neutral current and the $\rho$ parameter, large
Yukawa coupling of the top quark generates the mixing among its KK
modes and breaks the degeneracy: Unacceptably large contribution
to $\Delta \rho$ occurs. In order to satisfy the $\Delta\rho$
constraint, we relax the model by assigning a different bulk
fermion mass to SU(2) singlet bottom quark, and demonstrate that
there exists some limited parameter space where the $\Delta\rho$
constraint is satisfied. It is also shown that the current
measurement of Br($B \to X_s+\gamma)$ can be accommodated in this
modified model, and one sigma level precision constrains the
effective weak scale $k_{EW}$ such as $k_{EW} \gsim 3$~TeV for
$m_\psi/k=-0.4$, where $k_{EW}$ is the warp-suppressed AdS$_5$
curvature.
\end{abstract}

\maketitle

\section{Introduction}

Inspired by recent advances in string theories, some particle
physicists have resort to extra dimensions for the gauge hierarchy
problem of the standard model (SM). Arkani-Hamed, Dimopoulos and
Dvali (ADD) proposed that there exist $n$ large extra dimensions
with factorizable geometry {\cite{Antoniadis:1998ig}}: The observed
huge Planck scale $M_{\rm Pl}$ is attributed to the largeness of
the extra dimension volume $V_n$,
since $M_{\rm Pl}^2=M_S^{n+2} V_n$ with
$M_S$ being the fundamental string scale. The hierarchy problem is
resolved as      $M_S$ can be maintained around TeV. Later
Randall and Sundrum (RS) proposed another higher dimensional
scenario where the hierarchy problem is explained by
a geometrical exponential factor, based on two branes and a single extra dimension
with non-factorizable geometry {\cite{Randall:1999ee}}.
Apart from
the gauge hierarchy problem, TeV scale extra dimensions
accessible to the SM fields have drawn a lot of interest due to
various motivations such as gauge coupling
unification \cite{Dienes:1998vh}, new mechanisms for supersymmetry
breaking \cite{Antoniadis:1990ew}, the explanation of fermion mass
hierarchies \cite{Arkani-Hamed:1999dc}, and the presence of the
Higgs doublet as a composite of top
quarks \cite{Arkani-Hamed:2000hv}.

Of great significance and interest is that extra dimensional
models can leave distinct and rich phenomenological signatures at
future colliders. In the ADD scenario, due to the large volume of
extra dimensions, the SM fields should be confined to our brane.
Only the graviton propagates
in the bulk, to be observed as
Kaluza-Klein (KK) gravitons with almost continuous mass spectrum
\cite{ADD-ph}. In the models of TeV scale extra dimensions
accessible to the SM fields, KK excitations of the SM gauge (and
possibly fermion) fields with TeV scale masses would be observed
as resonances \cite{Uni-ph,Agashe:2001xt}. In the original RS
scenario, the SM fields are assumed to be confined to our brane.
Phenomenological signatures come from KK gravitons with
electroweak scale masses and couplings to matter, characterized by
$\Lambda_\pi$.
However, the small size of the RS-bulk allows
that the SM fields may also be in the bulk.
In Ref.\,\cite{Davoudiasl:1999tf}, it is demonstrated that placing
the SM gauge fields in the RS-bulk while confining the fermions to
our brane is strongly constrained by the current precision
electroweak data so that the lowest KK state of gauge boson should
be heavier than about 23~TeV. Then $\Lambda_\pi$ is pushed
up to about 100 TeV, which is disfavored as a solution of the gauge
hierarchy problem. If both the SM gauge and
fermion fields are in the bulk \cite{Davoudiasl:2000wi}, their
phenomenological signatures are very sensitive to the bulk fermion mass
$m_\psi$ which determines the KK mass spectrum of the bulk
fermions and their interactions with bulk gauge bosons.

In the early study of the RS-bulk SM,
Yukawa interactions with the Higgs field have been ignored
due to small quark masses compared to the KK mass scale.
Then a simple assumption of universal bulk fermion mass
suppresses their contributions to
flavor changing neutral current (FCNC) as well as
the $\rho$ parameter \cite{Davoudiasl:2000wi,Gherghetta:2000qt}.
This is because the common $m_\psi$ leads to the same KK mass spectra
for all the SM fermions.
The degeneracy of the up-type (and down-type) quark KK modes
operates the GIM cancellation \cite{GIM}
KK-level by level: With the
minimal flavor violation assumption that at the tree level the flavor mixing
comes only through the CKM matrix,
FCNC is suppressed as in the SM.
Since the KK modes of two constituents of SU(2)--doublet
have also the same mass,
their contribution to the $\rho$ parameter vanishes.
However, the Yukawa interaction with the Higgs field
mixes the fermion KK tower members, which can be substantial for
the top quark  \cite{Davoudiasl:2000wi,top-mixing}.
Recently it has been shown that
the large mixing in the top quark KK sector leads to unacceptably large
contribution to the $\rho$ parameter and raises $\Lambda_\pi$
above 100 TeV  \cite{Hewett:2002fe}.
To accommodate those electroweak precision data,
a `mixed' scenario is proposed with
the third generation fermions on the TeV brane
but the other generations in the bulk.
However, the first excited KK mode of
gauge bosons should be heavier than 11 TeV
due to the strong constraints from precision measurements:
It is hard to probe the new physics effects at LHC.
In addition, the obvious discrimination of fermions according to generation
may lead to potentially dangerous FCNC due to the absence
of GIM mechanism.

It must be worthwhile to keep the original framework
where all the SM particles except for the Higgs boson
are in the bulk, and to question other unsubstantiated  assumptions.
Then most of the rich phenomenologies of the RS-bulk SM
at future colliders can
remain still valid.
We propose to relax
the universal bulk fermion mass assumption.
As the simplest and minimal extension, and particularly to satisfy the $\rho$
constraint, we assign
a different bulk fermion mass $m_\psi'$ to
the SU(2)--singlet bottom quark field,
and see whether there exists some parameter space
to accommodate the $\rho$ and
$B\to X_s \gamma$ constraints.
We shall show that
this $m_\psi'$ allows some limited parameter space
where the degeneracy between the top and bottom quark KK modes is retained,
suppressing their contribution to the $\rho$ parameter.
Through a careful and thorough analysis,
the current constraint from the
$b\to s \gamma$ decay shall be also shown
consistent with the model.

This paper is organized as follows.
In Sec.~\ref{review}, we briefly review the
RS-bulk SM, focused on the
KK reduction and interactions of bulk gauge bosons and bulk fermions.
In particular, we point out a subtle point when
placing the SM fermions in the RS-bulk:
The fermion field contents should be doubled.
In Sec.~\ref{our-model},
we first discuss why the minimal model should be extended,
by estimating its large contribution to the $\rho$ parameter.
With a different bulk fermion mass for the SU(2)--singlet bottom quark field,
we  present
the full mass matrixes of the top and bottom quark KK modes,
including the Yukawa interactions with the Higgs field
confined to our brane.
Through explicit diagonalization of the top quark KK mass matrix,
we present generic features of the KK mode mixing.
In Sec.~\ref{rho-constraint},
we review the $\rho$ parameter and show there exists
some parameter space where the $\Delta\rho$ constraint is satisfied.
Section~\ref{bsg} deals with the detailed study of
the effects of the KK modes of $W$ gauge bosons and
up-type quarks
on the decay rate of $b \to s \gamma$.
Finally, Sec.~\ref{conclusion} includes summary and conclusions.

\section{Bulk SM in the RS scenario}
\label{review}

In the RS scenario, a single extra dimension has been introduced
with non-factorizable geometry, which is compactified on a
$S_1/Z_2$ orbifold of radius $r_c$  \cite{Randall:1999ee}.
Requiring four-dimensional Poincar\'{e} invariance, the RS
configuration has the following solution of the five-dimensional
Einstein's equation:
\begin{equation}
\label{metric}
d s^2=G_{MN} d x ^M d x ^N
= e^{-2 \sigma(\phi) } \eta_{\mu\nu} d x^\mu d x^\nu
+ r_c^2 d \phi^2,
\end{equation}
where the four-dimensional metric tensor is defined
by $\eta_{\mu\nu} ={\rm diag} (1,-1,-1,-1)$,
$\sigma(\phi) \ev k r_c |\phi|$,
$0 \leq |\phi| \leq \pi$,
and the five-dimensional curvature is $R_5=-20\, k^2$.
Upper-case Roman indices run over all the five dimensions
while the Greek indices over our four dimensions.
Two orbifold fixed points accommodate two three-branes,
the Planck brane at $\phi=0$ and
our TeV brane at $|\phi|=\pi$.
Due to the assignment of our brane at $|\phi|=\pi$,
any mass at the Planck scale appears
gravitationally red-shifted by the so-called warp factor of
$\es \equiv e^{-k r_c \pi}$.
If $k r_c \approx 12$,
the hierarchy problem can be resolved.
Four-dimensional Planck scale $M_{\rm Pl}$
is related with the fundamental string scale $M_5$
by $M_{\rm Pl}^2={M_5^3}(1-e^{-2kr_c \pi})/k$.
Now let us review the masses of the KK modes
of the bulk gauge and fermion fields, which are relevant to
the $\rho$ parameter and $b \to s \gamma$ decay.
The back-reaction of the bulk fields on the AdS$_5$ metric
is to be neglected.

\subsection{Gauge field KK spectrum}

For a massless SU(2) gauge field $A_M^a(x,\phi)$,
gauge invariant five-dimensional action is
\cite{Davoudiasl:1999tf,Pomarol:1999ad}
\beq
S_A =-\frac{1}{4}
\intfive \,G^{MK} G^{NL} F^a_{KL} F^a_{MN},
\eeq
where $F^a_{MN}=\partial_M  A^a_N - \partial _N A^a_M - g_5 \es^{abc}
A_M^b A_N^c ~(a,b,c=1,2,3)$.
By choosing the odd $Z_2$-parity for $A_5^a(x,\phi)$  \cite{Davoudiasl:1999tf}
and/or an appropriate gauge  \cite{Delgado:1998qr},
$A_5^a(x,\phi)$ decouples from the Lagrangian.
With the KK expansion of
\beq
\label{KK-A}
A^a_\mu(x,\phi) =\sum_{n=0}^\infty
A_\mu^{a(n)}(x) \frac{\chi_A^{(n)}(\phi)}{\sqrt{r_c}},
\eeq
we have a four-dimensional effective action
of massive KK gauge bosons as
\beq
S_A =\intfour \sum_{n=0}^\infty
\left[
-\frac{1}{4}
\eta^{\mu\kp}\eta^{\nu\lm} F^{a\n}_{\kp\lm} F^{a\n}_{\mu\nu}
-\frac{M_A^{(n)\;2}}{2} \eta^{\mu\nu} A^{a\n}_\mu A^{a\n}_\nu
\right]
,
\eeq
which is obtained by
\beq
\label{chi-A}
\chi^\n_A(\phi) =\frac{e^{\sg(\phi)}}{N_A^\n}
\left[
J_1(z_A^{(n)}(\phi)) +\al_A^{(n)} Y_1(z_A^{(n)}(\phi))
\right]\,.
\eeq
Here,    $z_A^{(n)}(\phi)$ and the normalization $N_A^\n$ are given by
\bea
z_A^{(n)}(\phi) &=& \frac{M_{A}^\n}{k} e^{\sg(\phi)}, \\ \no
N_A^\n  &=&
\frac{e^{k r_c \pi}}{\sqrt{k r_c}}
\left|
J_1(x_A^\n)
\right|,
\eea
where $x_A^\n \ev z_A^\n(\pi)=M_A^\n/k_{EW}$ with $k_{EW} \ev \es \,k$.
Note that the wave function $\chi^\n_A(\phi)$ satisfies the orthonormal condition
\beq
\label{otho-A}
\int_{-\pi}^\pi\; d \phi \;\chi_A^\m (\phi)\chi_A^\n(\phi) =\dt^{mn},
\eeq
which leads to $\chi^{(0)}_A ={1}/{\sqrt{2\pi}}$.
The continuity of $d \chi_A^{(n)}/d \phi $
at $\phi=0$ and $\phi=\pm \pi$ determines
the coefficient $\al_A^{(n)}$ to be
\beq
\al_A^{(n)} = -\frac{J_0({M^\n_{A}}/k)}{Y_0({M^\n_{A}}/k)}
\,,
\eeq
and     $x_A^{(n)}$
to be the roots of the following equation:
\beq
J_0(x_A^{(n)})+\al_A^\n Y_0(x_A^\n)=0.
\eeq
Numerically we have
$x_A^{(1)} \approx 2.45$,
$x_A^{(2)} \approx 5.57$,
$x_A^{(3)} \approx 8.70$,
and $x_A^{(4)} \approx 11.84$.

\subsection{Review of Grossman-Neubert fermion KK spectrum}
\label{GN}

Let us review in detail the KK solution of a bulk fermion
with arbitrary Dirac bulk mass
in the RS scenario  \cite{Gherghetta:2000qt,Grossman:1999ra,Chang:1999nh},
which causes a subtle problem
when discussing the bulk SM.
The five-dimensional action
of a Dirac fermion $\Psi$ with the bulk mass $m_\psi$
is
\beq\label{action}
   S = \int\!\mbox{d}^4x\!\int\!\mbox{d}\phi\,\sqrt{-G}
   \left\{ E_{\underline{A}}^A \left[ \frac{i}{2}\,\bar\Psi\gamma^{\underline{A}}
   (D_A-\overleftarrow{D_A})\Psi
   \right]  - m_\psi\,\mbox{sign}(\phi)\,\bar\Psi\Psi \right\} \,,
\eeq
where $D_A$ is the covariant derivative,
$\gamma^{\underline{A}}=(\gamma^\mu,i\gamma_5)$, and the
inverse vielbein
$E_{ \underline{A}}^A=\mbox{diag}(e^\sigma,e^\sigma,e^\sigma,e^\sigma,1/r_c)$.
The
underlined upper-case Roman indices describe
objects in the tangent
frame.
The contribution of the spin
connection $\omega_{\underline{B}\underline{C}A}$,
which vanishes
by including hermitian conjugates, is omitted.

Integration by parts leads to the action
\begin{eqnarray}
\label{Sdel}
   S = \int\!\mbox{d}^4x\!\int\!\mbox{d}\phi\,r_c\,&\bigg\{&
    e^{-3\sigma} \left( \bar\Psi_L\,i\rlap/\partial\,\Psi_L
    + \bar\Psi_R\,i\rlap/\partial\,\Psi_R \right)
    \nonumber\\
   &&\mbox{}- \frac{1}{2 r_c} \left[ \bar\Psi_L \left(
    e^{-4\sigma} \partial_\phi + \partial_\phi\,e^{-4\sigma}
    \right) \Psi_R - \bar\Psi_R \left(
    e^{-4\sigma} \partial_\phi + \partial_\phi\,e^{-4\sigma}
    \right) \Psi_L \right]
    \\ \no &&
    - e^{-4\sigma}\,m_\psi\,\mbox{sign}(\phi) \left(
    \bar\Psi_L \Psi_R + \bar\Psi_R \Psi_L \right)\bigg\}
     \,,
\end{eqnarray}
where we impose periodic boundary
conditions of $\Psi_{L,R}(x,\pi)
=\Psi_{L,R}(x,-\pi)$.
With the KK expansion of $\Psi$
\begin{equation}\label{KK}
   \Psi_{L,R}(x,\phi) = \sum_{n=0}^\infty
    \psi_{L,R}^\n(x)\,
   \frac{e^{2\sigma(\phi)}}{\sqrt{r_c}}\,\hat f_{L,R}^\n(\phi) \,,
\end{equation}
and the requirement of
\begin{eqnarray}
\label{cond}
   \int\limits_{-\pi}^\pi\!\mbox{d}\phi\,e^{\sigma(\phi)}
   \hat f^{\m*}_{L}(\phi)\,\hat f^\n_L(\phi)
   &=& \int\limits_{-\pi}^\pi\!\mbox{d}\phi\,e^{\sigma(\phi)}
    \hat f^{\m*}_{R}(\phi)\,\hat f^\n_R(\phi) = \delta^{mn} \,,
    \\ \label{LR}
   \left( \pm\frac{1}{r_c}\,\partial_\phi - m \right)
   \hat f^\n_{L,R}(\phi)
   &=& - M_{f}^{(n)}\,e^\sigma \hat f^\n_{R,L}(\phi) \,,
\end{eqnarray}
we have a tower of massive Dirac fermions
with the effective action
\begin{equation}\label{Sferm}
   S = \sum_{n=0}^\infty \int\!\mbox{d}^4x\,\Big\{
   \bar\psi^\n(x)\,i\rlap/\partial\,\psi^\n(x)
   - M_{f}^{(n)}\,\bar\psi^\n(x)\,\psi^\n(x) \Big\} \,.
\end{equation}

Note that $Z_2$-symmetric action constrains
$\bar\Psi\Psi=\bar\Psi_L\Psi_R+\bar\Psi_R\Psi_L$
($\Psi_{L,R}\equiv (1\mp \gamma_5)\Psi/2$) to be $Z_2$-odd, as can
be seen from the first term in Eq.\,(\ref{action}).
If $f_L^{(n)}$ is a $Z_2$-even function $\chi^{(n)}$ then
$f^\n_{R}$ should be a $Z_2$-odd function $\tau^{(n)}$ and vice
versa. With $\nu \ev m_\psi/k$ of order one, solutions are,
for $n \neq 0$,
\bea \chi^{(n)}(\phi) &=&
\frac{e^{\sg/2}}{N_\chi^\n} \left[ J_{1/2-\nu }(z^\n)+\be_\chi^\n
Y_{1/2-\nu}(z^\n) \right], \\ \no \tau^{(n)}(\phi) &=&
\frac{e^{\sg/2}}{N_\tau^\n} \left[ J_{1/2+\nu} (z^\n)+\be_\tau^\n
Y_{1/2+\nu}(z^\n) \right], \eea and for $n=0$ \beq \label{chi-0}
\chi^{(0)} (\phi)=\frac{e^{\nu\sg(\phi)}}{N_\chi^{(0)}},~~
\tau^{(0)} (\phi)=0.
\eeq

With the even $Z_2$-parity of $\chi^\n$ and the odd parity of $\tau^\n$,
Eq.\,(\ref{LR}) yields boundary conditions
\beq
\label{BC}
0=
\left(
\frac{d }{d\phi}-m r_c
\right)\left.\chi^\n \right|_{\phi=0,\pi}=
\left. \tau^\n \right|_{\phi=0,\pi}
\,,
\eeq
which determine the coefficients $\beta^\n_{\chi,\tau}$
and the normalization constants $N^\n_{\chi,\tau}$,
as well as the KK fermion mass $M_f^\n \ev x_f^\n k_{EW}$
with $x_f^\n \equiv z^\n_f(\pi)$.
We present only $Z_2$-even part relevant for later discussion:
\bea
\be_\chi^\n &=& -\frac{
J_{-(\nu+1/2)} (M_f^\n/k)
}{Y_{-(\nu+1/2)} (M_f^\n/k)
}\,,
\\ \no
N_\chi^{(0)} &=&
\frac{1}{\es^{(\nu+1/2)}}
\sqrt{
\frac{2}{k r_c }\left|\frac{1-\es^{1+2\nu}}{1+2\nu}\right|}
\,\,, \\ \no
N_\chi^\n&=&
\frac{e^{kr_c\pi}}{x_f^\n \sqrt{k r_c}}
\sqrt{ \left.
z_\chi^{\n 2}\left\{J_{{1\over 2}-\nu}(z_\chi^\n)
+\be_\chi^\n Y_{{1\over 2}-\nu}(z_\chi^\n)\right\}^2
\right|_{\phi=0}^{\phi=\pi}
}\,\,.
\eea
And    $x_f^\n$ is the solution of
\beq
J_{-(\nu+1/2)}(x_f^\n) +
\be_\chi^\n Y_{-(\nu+1/2)}(x_f^\n)=0
\,,
\eeq
which is the same as that of the right-handed one.
We refer to Ref.\,\cite{Davoudiasl:2000wi}
for the corresponding expressions of the $Z_2$-odd part.

Discussions on the physical and
phenomenological implications of the parameter $\nu$
are in order here.
Note that the canonically re-scaled zero mode of $Z_2$-even bulk fermion
is
proportional to $e^{ (1/2 +\nu)k r_c |\phi|}$
(see the first line of Eq.\,(\ref{Sdel}) with
Eqs.\,(\ref{KK}) and (\ref{chi-0})).
For $\nu \ll -1/2$ the fermion bulk wave functions
are localized toward  the Planck
brane:
The magnitudes of its gauge couplings with KK gauge bosons
are quite small.
In Refs.~ \cite{Gherghetta:2000qt,Davoudiasl:2000my},
it has been numerically demonstrated
that for $\nu \lsim -0.5$ the couplings are too small
to be probed at high energy colliders.
For $\nu \gg 1$
the SM fermions become localized closer to the TeV brane:
The model approaches the RS model with the gauge fields only
in the bulk, which is phenomenologically disfavored
due to unreasonably large $M_A^{(1)}$  \cite{Davoudiasl:1999tf}.
To be specific, if $\nu \gsim -0.3$,
the large contribution to the precision electroweak data
pushes $M_A^{(1)}$
up to about 6~TeV,
beyond the direct production at any planned collider.
In what follows, therefore, we consider the parameter space of $\nu$
between $-0.5$  and $-0.3$.

\subsection{Accommodating the SM fermion sector}

In order to
place the SM fermions in the AdS$_5$ bulk,
the fermion field contents should be doubled.
In the SM, a fermion field with left-handed chirality and
that with right-handed chirality belong to
different representations of a gauge group:
For example, the left-handed up quark ($u_L$)
and the left-handed down quark ($d_L$)
form an SU(2)--doublet,
while the right-handed up and down quarks ($u_R$ and $d_R$)
are two SU(2)--singlet fields.
 $u_L$ and $u_R$ are related by the Dirac mass term,
Yukawa coupling with the Higgs field.
Two Dirac fermion fields, $u$ and $d$,
are enough to describe each generation in the quark sector.
In the RS-bulk SM,
a fermion which belongs to a specific representation of a gauge group
should possess both left- and right-handed chiralities.
If only one chiral state  (e.g.,  $\Psi_L$) exists,
the second line in Eq.\,(\ref{Sdel})
as well as the bulk mass term
vanish: Non-trivial solution of the $\widehat{f}^\n_{L,R}(\phi)$
cannot be obtained.
For each generation,
we introduce four five-dimensional Dirac fields, an SU(2)--doublet
fermion field $Q=(q_u, q_d)^T$
and two SU(2)--singlet fermion fields,
$u$ and $d$,
with weak hypercharges
$Y=1/6$, $2/3$, and $-1/3$ respectively.

Since the SM fermion should correspond to the KK zero mode,
we assign $Z_2$-even wave function $\chi^\n$
to the left-handed SU(2)--doublet
and the right-handed SU(2)--singlet such that
\bea
\label{KK-f}
   Q(x,\phi) &=&Q_L + Q_R= \sum_n
   \frac{e^{2\sigma(\phi)}}{\sqrt{r_c}}
   \left[
   Q_L^\n(x) \chi^\n(\phi)
   +
   Q_R^\n(x) \tau^\n(\phi)
   \right],\\ \no
   u(x,\phi)&=&u_L + u_R=\sum_n
   \frac{e^{2\sigma(\phi)}}{\sqrt{r_c}}
   \left[
   u_L^\n(x) \tau^\n(\phi)
   +
   u_R^\n(x) \chi^\n(\phi)
   \right],
\eea
and  $d(x,\phi)$ has the same KK decomposition as  $u(x,\phi)$.
The charged current interactions,
mediated by the bulk $W$ boson,
connect $q_u$ and $q_d$:
\bea
S_{f\bar{f}'W^\pm}
&=& \intfive e^\sg \frac{g_5}{\sqrt{2}}
\left[
\bar{q}_u \rlap/W^+ q_d +h.c.
\right]
\\ \label{CC} &=&
\intfour
\frac{g}{\sqrt{2}}
\sum_{l=0}^\infty
\left[ \sum _{n,m=0}^\infty
\bar{q}_{uL}^\n \rlap/W^{+(l)} q_{dL}^{(m)}
\left\{
       C_{nml}^{\bar{f}f' W}
\right\}
\right.
\\ \no &&~~~
+ \left.\sum _{n,m=1}^\infty
\bar{q}_{uR}^\n \rlap/W^{+(l)} q_{dR}^{(m)}
\left\{
\sqrt{2\pi}\int_{-\pi}^\pi d\phi \,e^\sg \tau^\n \tau^{(m)}\chi_A^{(l)}
\right\}\right]+h.c.,
\eea
where $g ={g_5}/{\sqrt{2\pi r_c}} $,
and the KK expansion in Eqs.\,(\ref{KK-A}) and
(\ref{KK-f}) have been substituted.
$C_{nml}^{\bar{f}f' W}$ denotes the coupling of the $m$-th and
the $n$-th fermion states to the $l$-th $W$ boson
in the unit of the SM coupling. It is defined by
\beq
\label{C}
C_{nml}^{\bar{f}f' W}
=\sqrt{2\pi}
\int_{-\pi}^\pi d\phi \, e^\sg \chi^\n(\phi)
\chi^{(m)}(\phi)\chi_A^{(l)}(\phi).
\eeq

\section{Phenomenologically Viable Minimally Extended Model}
\label{our-model}
\subsection{Relaxing the universal bulk fermion mass assumption}

Now every SM fermion possesses its KK tower.
We remind the reader that in the RS background the fermion KK mass spectrum
is determined by the bulk fermion mass $m_\psi$.
A simple assumption of universal bulk fermion mass
results in the same KK mass spectrum for all the SM fermions.
The exact degeneracy between the KK masses for the $T_3=\pm 1/2$ fermions
cancels their contribution to the $\rho$ parameter.
In addition,
the degeneracy among the KK masses of up-type quarks (and down type quarks)
allows the GIM cancellation to occur KK-level by level.

However, there is another mass source,
Yukawa interaction.
This Yukawa mass relates SU(2)--doublet with singlet
(e.g., $m_{Y}\overline{q}_{uL}u_{R}$).
It is to be compared to the KK mass which is the coupling of
$\overline{q}_{uL} q_{uR}$ and $\overline{u}_{L} u_{R}$.
This difference results in mixing among the fermion KK modes.
Since quark masses are much smaller than the KK mass scale,
this mixing effect has been neglected  in the early study.
One exception is the top quark.
Its heavy mass yields substantial mixing among
top quark KK modes.
The mass shift of the top quark KK mode from the rest up-type quark KK modes
invalidates the GIM cancellation;
FCNC becomes inevitable.
More severe problem happens if precision measurements
are taken into account.
In particular, the $\rho$ parameter becomes dangerous
since it does not follow the decoupling theorem in
the sense that its quantum correction
increases with the squared mass difference between the $T=1/2$ and
$T=-1/2$ fermions.
As each top quark KK mass deviates from the
corresponding bottom quark KK mass,
their additive contribution yields disastrous and unacceptable
value of $\Delta \rho$.
The problem does not ameliorate but worsens as we add
more and more KK states \cite{Hewett:2002fe}.

In Ref.\,\cite{Hewett:2002fe},
a `mixed' scenario has been proposed such that
the third generation fermions are confined on the TeV brane
while the other two generations propagate in the bulk.
This construction itself is interesting with the additional
attractive feature of its natural explanation for
the observed $m_c/m_t$ and $m_s/m_b$ hierarchies.
However,
constraints from current precision measurements
are rather strong that the first excited KK mode of
gauge bosons is about 11 TeV, not leading to any new physics signatures
at LHC.
In addition,
the obvious discrimination of fermions according to generation
may lead to potentially dangerous FCNC due to the absence
of GIM mechanism:
Without a detailed analysis of FCNC effects,
Ref.\,\cite{Hewett:2002fe} claimed that the estimated size of KK $Z$ boson
exchange effects on the $K-\overline{K}$ mixing is
within the uncertainty of the SM results.

Instead, here we keep the original set-up but relax
the unsubstantiated assumption of the universal bulk fermion mass.
For the simplest extension, we assume that
the SU(2)--singlet bottom quark field has different bulk fermion mass $m_\psi'$,
and see whether this introduction of another parameter
can accommodate the $\Delta\rho$ constraint without a new hierarchy.
Then the five dimensional action for the
third generation quarks becomes
\bea \label{actionSM}
   S = \int\!\mbox{d}^4x\!\int\!\mbox{d}\phi\,\sqrt{-G}
   &&\left[ E_a^A \left(
   i\,\bar{Q}\gamma^a \D_A Q
 + i\,\bar{t}\gamma^a \D_A t
 +i\,\bar{b}\gamma^a \D_A b
   \right)  \right.\\ \no
   && \left.
    - \mbox{sign}(\phi)\left(
    m_\psi\left\{ \bar{Q}Q + \bar{t}t\right\}
    +m_\psi'\,\bar{b}b \right)\right] \,.
\eea
The
compactification of the extra dimension leads to the KK mass terms such that
\beq
\label{KK-mass}
\Lg =- \sum_{n=1}^\infty k_{EW} \left[
x_f^\n(\nu)\left\{ \bar{q}^\n_{t L} q_{tR}^\n +\bar{q}^\n_{b L}
q_{b R}^\n
 + \bar{t}^\n_L
t_R^\n  \right\}+{x_f^\n}(\nu\,') \bar{b}^\n_L b_R^\n \right] + h.c.,
\eeq
where we introduced additional dimensionless parameter $\nu\,'=m_\psi'/k$.
As explicitly shown in Eq.\,(\ref{KK-mass}),
the KK masses of fermions depend on the
bulk fermion masses, $m_\psi$ and $ m_\psi'$.

\subsection{Mass matrix of the KK modes for the top and bottom quarks}

In addition to the KK masses,
an observer on the TeV brane
has another source for the fermion mass,
Yukawa coupling with the Higgs boson.
The Higgs mechanism should operate here so that
KK zero modes of the bulk gauge boson and fermion,
which correspond to the SM particles,
acquire masses.
However,
the simplest case with
the Higgs boson in the bulk results in some unsatisfactory consequences.
First,
the lowest mass eigenvalue of the gauge boson,
proportional to the bulk mass of the Higgs field,
has no suppression by the warp factor  \cite{Chang:1999nh}.
The Higgs bulk mass should be much smaller
than the Planck mass scale.
The gauge hierarchy problem has recurred.
Second, the bulk Higgs mechanism cannot retain
the correct SM gauge couplings of the photon, $W$ and $Z$ bosons
if the SM fermions are in the bulk;
if the fermions are on the wall,
the SM mass relationship of the $W$ and $Z$ bosons
is broken \cite{Davoudiasl:2000wi}.
It is concluded that at least one Higgs
field must be confined to the TeV brane.

The five-dimensional action for Yukawa interaction with the
confined Higgs field is
\bea \label{KK-f-mass}
S_{ffH}&=&
-\intfive \left[  \frac{\lm_5^b}{k} \,{\overline{Q}}(x,\phi) \cdot H(x) b
(x,\phi) \right.
\\ \no
&&\qquad \left. + \frac{\lm_5^t}{k} \,\es^{ab} \overline{Q}(x,\phi)_a \cdot  H(x)_b t(x,\phi)
+h.c. \right] \dt(\phi-\pi) \,,
\eea
where $\lm_5^{b,t}$ is the
five-dimensional Yukawa coupling.
Spontaneous symmetry breaking shifts the Higgs
field as $H^0 \to v_5 +H^{'0}$ with a VEV of the order Planck
scale.
Then the four-dimensional effective Lagrangian becomes
\bea
\label{Yukawa}
\Lg_{eff} &=&\frac{\lm_t v}{\sqrt{2}} \left(
\bar{q}^{(0)}_{tL} +\hat{\chi}_1 \bar{q}^{(1)}_{tL} +\cdots
\right) \left( t^{(0)}_R +\hat{\chi}_1 t^{(1)}_R+\cdots \right)
\\ \no &&+
\frac{\lm_b v}{\sqrt{2}} \left( \bar{q}^{(0)}_{bL} +\hat{\chi}_1
\bar{q}^{(1)}_{bL} +\cdots \right) \left( b^{(0)}_R
+\hat{\chi}'_1 b^{(1)}_R+\cdots \right) \,,
\eea
where $\lm_{t,b}
={\lm_5^{t,b} }(1+2\nu)/{2}(1-\es^{1+2\nu})$, $v =\es v_5$,
$\hat{\chi}_n \equiv \chi^\n(\pi,\nu)/\chi^{(0)}(\nu)$, and
$\hat{\chi}'_n$ are with $\nu\,'$.
Note that the right-handed SU(2)--doublet $q_{tR}^\n$ and
the left-handed SU(2)--singlet $t_L^\n$
do not appear in the Yukawa term due to the $Z_2$-odd boundary conditions
in Eq.\,(\ref{BC}).

For definite presentation, we introduce the number of the KK
states, $n_\infty$, which is in principle infinity.
Equations (\ref{KK-mass}) and (\ref{Yukawa}) imply the mass terms
of the KK modes as follows:
\beq \label{top-mass}
\Lg_{mass} = - \left(\bar{t}_R^{(0)} \bar{t}_R^{(1)} \cdots |\,
\bar{q}^{(1)}_{tR} \cdots \right) {\mathcal M}_t
\left(\begin{array}{c}
  q_{tL}^{(0)} \\
  q_{tL}^{(1)} \\
  \vdots \\ \hline
   t_L^{(1)}\\
  \vdots
\end{array}
\right)
- \left(\bar{b}_R^{(0)} \bar{b}_R^{(1)}
\cdots |\, \bar{q}^{(1)}_{bR} \cdots \right) {\mathcal M}_b
\left(\begin{array}{c}
  q_{bL}^{(0)} \\
  q_{bL}^{(1)} \\
  \vdots \\ \hline
   b_L^{(1)}\\
  \vdots
\end{array}
\right)\,,
\eeq
where the $(2 n_\infty+1)\times (2 n_\infty+1)$ matrix
${\mathcal M}_{t,b}$ are defined by
\beq
\label{M}
{\mathcal M}_t=
\left(
\begin{array}{cc}
 {\mathcal M}_Y^t & {\mathcal M}_{KK}^{t} \\
{\mathcal M}_{KK}^{q_t} & 0
\end{array}
\right),\quad
{\mathcal M}_b=
\left(
\begin{array}{cc}
 {\mathcal M}_Y^b & {\mathcal M}_{KK}^{b} \\
{\mathcal M}_{KK}^{q_b} & 0
\end{array}
\right).
\eeq
The $(n_\infty+1)\times (n_\infty+1)$ matrix
${\mathcal M}_Y^{t,b}$ is from Yukawa mass terms, and the $(n_\infty+1)
\times n_\infty$ matrix ${\mathcal M}_{KK}$ from KK masses. They
are given by
\bea
\label{MY}
{\mathcal M}_Y^t &=& m_{t,0} \left(
\begin{array}{cccc}
  1 & \hat{\chi}_1  & \hat{\chi}_2 & \cdots \\
  \hat{\chi}_1  &\hat{\chi}_1^2
  & \hat{\chi}_1 \hat{\chi}_2  & \cdots \\
  \hat{\chi}_2  & \hat{\chi}_2 \hat{\chi}_1
  & \hat{\chi}_2 ^2  & \cdots\\
  \vdots & \vdots & \vdots & 
\end{array}
\right) ,\quad
{\mathcal M}_Y^b = m_{b,0}
\left(
\begin{array}{cccc}
  1 & \hat{\chi}_1  & \hat{\chi}_2 & \cdots \\
  \hat{\chi}'_1  &\hat{\chi}'_1\hat{\chi}_1
  & \hat{\chi}'_1 \hat{\chi}_2  & \cdots \\
  \hat{\chi}'_2  & \hat{\chi}'_2 \hat{\chi}_1
  & \hat{\chi}'_2\hat{\chi}_2  & \cdots\\
  \vdots & \vdots & \vdots & 
\end{array}
\right),\no  \\
{\mathcal M}_{KK}^{q_t} &=&
k_{EW} \left(
\begin{array}{cccc}
  0 & x_f^{(1)} & 0 & \cdots \\
  0 & 0 & x_f^{(2)} &\cdots \\
  0 & 0 &  0   & \cdots \\
  \vdots & \vdots & \vdots & 
\end{array}
\right), \qquad
{\mathcal M}_{KK}^{b} =
k_{EW}
\left(
\begin{array}{ccc}
  0 & 0 & \cdots \\
  x_f^{(1)}(\nu\,') & 0 & \cdots \\
  0 &  x_f^{(2)}(\nu\,') & \cdots \\
  \vdots & \vdots & 
\end{array}
\right),
\eea
where $m_{t(b),0}=\lm_{t(b)} v/\sqrt{2}$ of electroweak
scale, and
${\mathcal M}_{KK}^{q_b}={\mathcal M}_{KK}^{t}={\mathcal M}_{KK}^{q_t}$.
Note that the bottom-right blocks of ${\mathcal M}_{t,b}$
vanish since there are no Yukawa couplings of $Z_2$-odd bulk fermions.

\subsection{Diagonalization of up-type quark KK mass matrix}

Let us discuss the diagonalization of the ${\mathcal M}_{u,c,t}$ in detail.
This shows generic features of KK mode mixing in the RS scenario,
as well as being
relevant for the $b\to s\gamma$ decay.
Since even in the modified model
all the up-type quark fields have the same bulk fermion mass,
$\M_{u(c)}$ is the same as
$\M_{t}$
with the replacement of $m_{t,0}$ by $m_{u(c),0}$,
collectively denoted by $\M_q$.
A real and symmetric matrix ${\mathcal M}_q$
is diagonalized by an orthogonal matrix ${\mathcal N}$:
\beq
{\rm diag} (\eta_0, \eta_1, \cdots)
{\mathcal N}{\mathcal M}_q
{\mathcal N}^T =
{\rm diag}(m_q, M_1, M_2,\cdots),
\eeq
where $\eta_{j}=\pm 1$
is introduced for the positive-definite mass.
 $q_{uL}^\n$ and $u_L^\n$
form $(2\,n_\infty+1)$ left-handed mass eigenstates ${u'_L}^{(\underline{j})}$
\beq
\left(\begin{array}{c}
  {u'_L}^{(0)} \\
  {u'_L}^{(1)} \\
  {u'_L}^{(2)} \\
  \vdots
\end{array}
\right)
= {\mathcal N}
\left(
\begin{array}{c}
  q_{uL}^{(0)} \\
  q_{uL}^{(1)} \\
  \vdots \\ \hline
  u_{L}^{(1)} \\
  \vdots
\end{array}
\right)
\,.
\eeq
For example, $q^{(n)}_{uL}$
is a mixture of KK mass eigenstates ${u'}_L^{(\underline{j})}$
\beq
q^{(n)}_{uL} =\sum _{\underline{j}=0}^{2 n_\infty}
{\mathcal N}_{(\underline{j},\,n)} {u_L'}^{(\underline{j})}, \quad (n=0, \cdots, n_C)
\,.
\eeq
In what follows, the underlined index runs from zero to $2\, n_\infty$.
In terms of mass eigenstates,
the four-dimensional effective Lagrangian of the charged current,
relevant for the FCNC, is
\beq
\Lg = \frac{g}{\sqrt{2}} V_{tb}
\left(
\sum_{m=0}^{n_\infty}
\C_{0ml} {\mathcal N}_{(\underline{j},\,m)}
\right)
\bar{b}_L^{(0)} \gm^\mu t_L^{'(\underline{j})} W^{(l)}_\mu + h.c.
.
\eeq

Now let us discuss the diagonalization of ${\mathcal M}_q$.
For light quarks ($m_{q,0}=0$),
 ${\mathcal M}_q$ can be analytically diagonalized.
Mass eigenvalues are
\beq
\label{mass0}
m_q^{\,(0)}=0,\quad
 M_n^{\,(0)}=
 M_{n_\infty+n}^{\,(0)}= x_f^{(n)} k_{EW},
\eeq
which are obtained by the orthonormal matrix ${\mathcal N}^{(0)}$:
\bea
\label{N0}
{\mathcal N}_{(0,\,0)}^{\,(0)} =1,
\quad
{\mathcal N}_{(0,\,\underline{n})}^{\,(0)} =
{\mathcal N}_{(\underline{n},\,0)}^{\,(0)} =0, \quad
{\mathcal N}_{(n,\,m)}^{\,(0)} =-{\mathcal N}_{(n_\infty+n,\,m)}^{\,(0)}=
\frac{\dt_{n m}}{\sqrt{2}}.
\eea

\begin{figure}
\includegraphics[height=10cm,angle=-90]{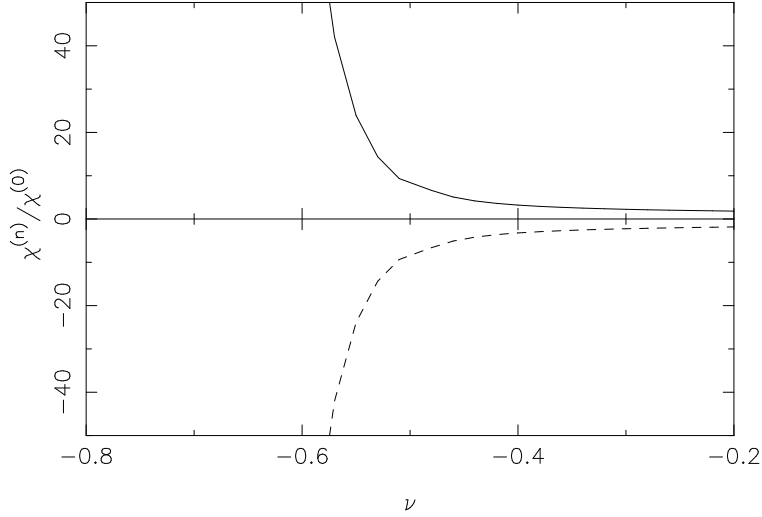}
\caption{\label{chi} $\hat{\chi}^{(n)} \equiv
\chi^{(n)}/\chi^{(0)}$ as a function of $\nu$ with $k r_c=11.5$.
The solid line is for odd modes ($n=1,3,5,...$) and the dotted
line for even modes ($n=2,4,6,...$). }
\end{figure}
For top quark with non-negligible $m_{t,0}$,
the diagonalization of ${\mathcal M}_q$ is not
trivial.
Since  $m_{t,0}$ is approximately the top quark mass
of 175 GeV while the KK fermion masses
are of the order TeV,
the diagonalization can be made perturbatively
unless  $\hat\chi^2_n$'s are much larger than unity.
In Fig.\,\ref{chi}, we present the values of $\hat{\chi}_n$
which are very sensitive to  $\nu$:
The solid line is for $\hat{\chi}_{2n-1}$ and
the dashed line is for $\hat{\chi}_{2n}$ $(n=1,2,\cdots)$.
Note that $\hat{\chi}^{(n)}$ does not depend on $k_{EW}$.
For $\nu \lsim -0.4$,
 $|\hat{\chi}_n|$ rapidly increases so that the elements
of  ${\mathcal M }_Y$
can be compatible or even larger than some elements of  ${\mathcal M}_{KK}$:
Diagonalization is to be made only numerically.

For $\nu \gsim -0.3$,
 $m_{t,0}/k_{EW}$ is a good perturbation parameter, denoted by $\delta$.
To leading order in $\dt $,
the mass eigenvalues are
\beq
\label{mass}
m_q=m_0,\quad
 M_n =  k_{EW}\left(x_f^{(n)}+\frac{\hat{\chi}_n^2}{2}\,\dt \right),
 \quad
 M_{n_\infty+n}= k_{EW}\left(x_f^{(n)}-\frac{\hat{\chi}_n^2}{2}\,\dt \right),
\eeq
and the elements of the orthonormal matrix ${\mathcal N}$ are
parameterized by
\beq
{\mathcal N}_{(\underline{n},\,\underline{m})}
\equiv {\mathcal N}_{(\underline{n},\,\underline{m})}^{\,(0)}+
 {\mathcal N}^{\,(1)}_{(\underline{n},\,\underline{m})} \frac{\dt }{\sqrt{2}}
,
\eeq
where
\bea\label{N}
{\mathcal N}^{\,(1)}_{(0,\,0)} &\simeq& 0, \quad
{\mathcal N}^{\,(1)}_{(0,\,n)} \simeq 0, \quad
{{\mathcal N}^{\,(1)}}_{(n,\,0)}
\simeq {\mathcal N}^{\,(1)}_{(n_\infty+n,\,0)}
\simeq \frac{\hat{\chi}^2_n}{x_f^{(n)}},
\\ \no
{{\mathcal N}^{\,(1)}}_{(n ,\, n)}
&\simeq& {\mathcal N}^{\,(1)}_{(n_\infty+n,\,n)} \simeq
\frac{\hat{\chi}^2_n}{4 x_f^{(n)}}\, ,
\\ \no
{{\mathcal N}^{\,(1)}}_{(n ,\, m)} &\simeq&
{{\mathcal N}^{\,(1)}}_{(n_\infty+n ,\, m)} \simeq
\frac{x_f^{(n)} \hat{\chi}_n \hat{\chi}_m}
{\sqrt{2}({x_f^{(n)\,2}}-{x_f^{(m)\,2}})}\,
\quad \mbox{for}~~(n\neq m)\,.
\eea

\section{Constraints from the $\rho$ parameter}
\label{rho-constraint}

The $\rho$ parameter
is defined by the difference between the $W$ and $Z$ boson self-energy
functions re-scaled by each mass:
\beq
\rho =
\frac{\Pi_W(q^2=0)}{m_W^2}
-\frac{\Pi_Z(q^2=0)}{m_Z^2}
\,.
\eeq
It
has been known to
play a special role among
precision measurements
since it is sensitive to heavy fermions beyond SM:
Its quantum correction
increases with the mass difference between two constituent fermions
of SU(2)--doublet.
A SU(2)--doublet $(f_u,f_d)^T$ yields additional contribution to $\rho$:
\begin{equation}
\Delta\rho=\frac{N_c G_F}{8\sqrt{2}\pi^2}\Delta m^2,
\end{equation}
where $N_c$ is the number of colors and $\Delta m^2$ is given by
\begin{equation}
\Delta m^2 \equiv m_u^2+m_d^2-\frac{4 m_u^2
m_d^2}{m_u^2-m_d^2}\ln\frac{m_u}{m_d} .
\end{equation}
We denote masses of $f_u$ and $f_d$ fermions as $m_u$ and $m_d$, respectively.
As can be seen from the $\Delta m^2$
in two limiting cases,
\begin{equation}
\Delta m^2 \simeq
\left\{
\begin{array}{ll}
  m_u^2 & \quad \mbox{  for  } m_u\gg m_d \\
  (m_u-m_d)^2 & \quad \mbox{   for  } m_u \simeq m_d
\end{array} \right.  ,
\end{equation}
the contribution of any degenerate SU(2)--doublet to $\Delta\rho$
vanishes.
With the Higgs mass below 1 TeV,
the current electroweak precision data constrain
$\Delta\rho<2\times 10^{-3}$  at 95\% CL
(with $\Delta\rho \equiv \rho-\rho_{\rm SM}$),
which restricts new SU(2)--doublet
fermion mass spectrum to satisfy  \cite{PDG}
\begin{equation}\label{rho}
\sum_i \Delta m_i^2 \le (115 \,\mbox{GeV})^2 .
\end{equation}

\begin{figure}
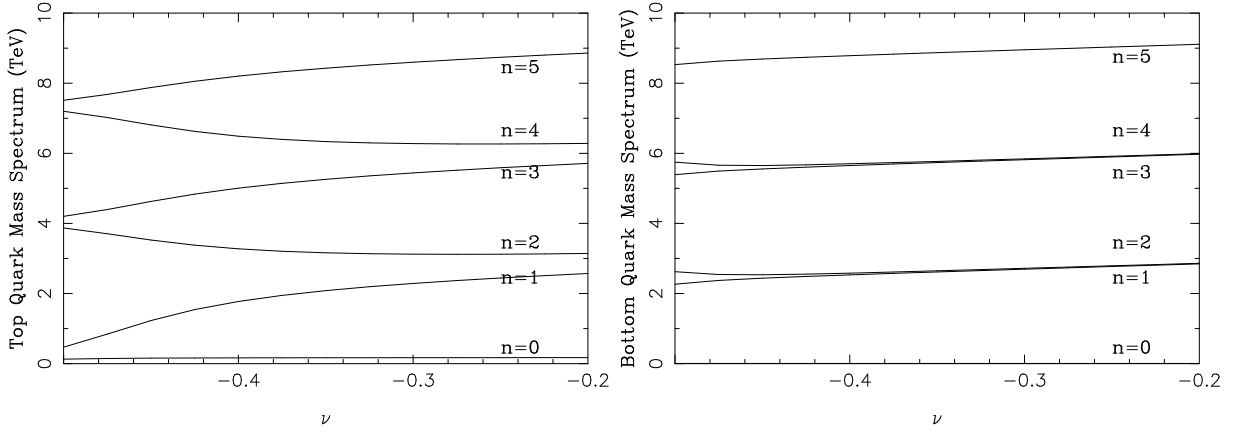

\includegraphics[height=8cm,angle=-90]{fig2a}
\includegraphics[height=8cm,angle=-90]{fig2b}
\caption{\label{fig1} The KK masses for (a) the top  and (b) bottom quarks
when $\nu=\nu'$ and $k_{EW}=1$ TeV.
Here we have employed a full numerical diagonalization.
}
\end{figure}
First let us demonstrate that the RS-bulk SM with the universal bulk fermion
mass assumption ($\nu=\nu'$)
cannot satisfy this $\Delta\rho$ constraint.
Now that every SM fermion has its KK tower,
we have various vacuum polarization graphs mediated by
the fermion KK modes for the $W$ and $Z$ self energies.
Since involved couplings incorporate the zero-mode
of gauge boson KK modes,
the simple relation of $C^{ff'A}_{nm0}=\dt_{nm}$
allows us to ignore the $\Delta m^2$ between different KK levels, to leading order.
For degenerate top and bottom quark KK modes,
i.e., for $|M_t^\n-M_b^\n| \ll M_t^\n$,
we have
\beq
\Delta m^2 \approx \sum_n (M_t^\n-M_b^\n)^2
\,.
\eeq
Figure \ref{fig1} presents the KK masses for (a) the top and (b) bottom quarks.
Here we set $kr_c=11.5$, $k_{EW}=1$ TeV and the zero mode
top quark mass to be 175 GeV and the zero mode bottom quark mass
$4.5$ GeV.
As $\nu$ increases, the mass of an odd mode ($n=1,3,\cdots$) approaches
that of the corresponding even mode ($n=2,4,\cdots$).
For the bottom quark case,
two KK modes are almost degenerate in most of the relevant parameter space.
This is because of the small $m_{b,0}$ compared to $k_{EW}$.
These different mass spectra for the top and bottom quark KK modes
lead to dangerous contributions to $\Delta\rho$.
\begin{figure}
\includegraphics[height=9cm,angle=-90]{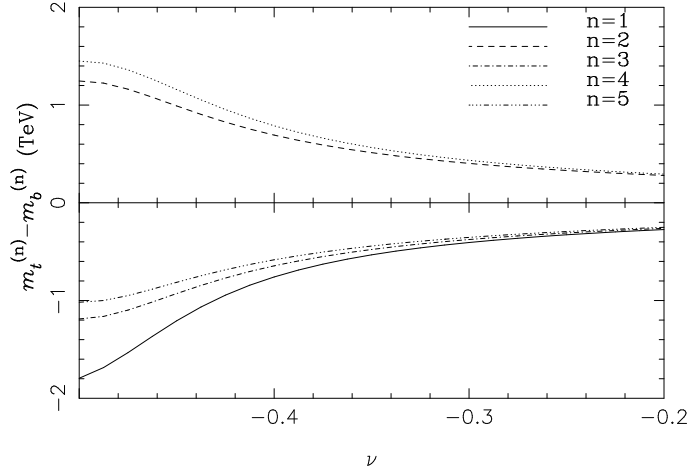}
\caption{\label{deltamuni} The KK mass difference
between the top and bottom quarks
when $\nu=\nu'$ and $k_{EW}=1$ TeV.
}
\end{figure}
In Fig.\,\ref{deltamuni}, we show
top and bottom KK mass differences up to the fifth KK excitation
states as a function of parameter $\nu$.
Even though the $\Delta m$ decreases with $\nu$,
it is too large to satisfy Eq.\,(\ref{rho}).

\begin{figure}
\includegraphics[height=9cm,angle=-90]{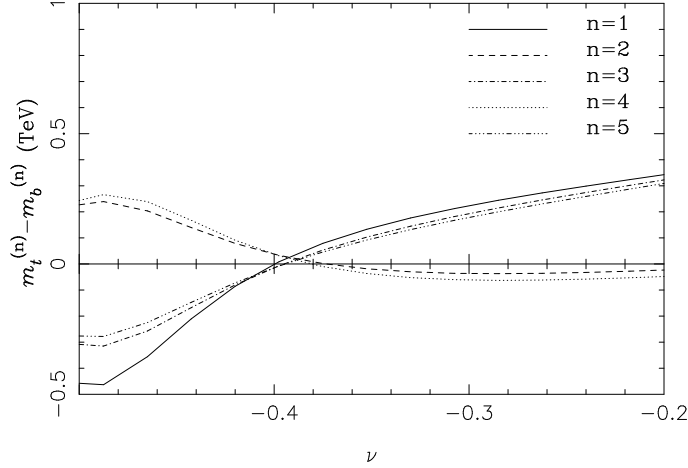}
\caption{\label{deltam} The KK mass difference
between the top and bottom quarks
when $\nu'=-0.6$ and $k_{EW}=1$ TeV.
}
\end{figure}

Now  let us allow $\nu \neq\nu'$
and see whether there exists a parameter space
where the degeneracy between the top and bottom quark KK modes
is retained.
We need large mixing among the bottom quark KK modes
so that the rapid increase of $M_b^{(1)}$ in Fig.\,\ref{fig1}
can be slowed down.
As discussed before,
large mixing occurs with large negative $\nu'$.
In Fig.\,\ref{deltam}, we
shows, with the fixed $\nu'=-0.6$, the
top-bottom quark mass differences for the first five excited
modes. We find that for example the $\nu=-0.39$ case with $\nu\,'=-0.6$
gives vanishing
mass differences below 20 GeV for the first five KK excited
states.
A remarkable point is that
the $\Delta m$ decreases for higher KK
modes; the contribution of higher KK modes becomes less important.

\section{Constraints from $B\to X_s \gamma$ decay}
\label{bsg}
It is well known that any kind of new physics beyond the SM is
significantly constrained by flavor changing neutral current
(FCNC), which the SM predicts to be suppressed at one loop level
by light quark masses relative to $M_W$ and by small CKM mixing
between the third and first two generations.
In particular,
the rare decay of $B \to X_s \gamma$
has been extensively studied within and beyond the SM
 \cite{Greub:1997hf}.
Moreover, this decay mode is sensitive to the top quark sector,
appropriate to probe any new physics related with top quark.
The inclusive decay $B \to X_s \gamma$ is approximated
by the partonic decay $b \to s \gm$ with the following equality:
\beq
\frac{\Gamma(B \to X_s \gm)}{\Gamma (B \to X_c e \bar{\nu}_e)}
\simeq
\frac{\Gamma(b \to s \gm)}{\Gamma (b \to c e \bar{\nu}_e)}
\ev R_{\rm quark}
\,.
\eeq
With the NLO QCD corrections,
 $R_{\rm quark}$ is  \cite{Buras:1998ra}
\beq\label{R}
R_{\rm quark} =\frac{ \lm_t^2}{|V_{cb}|^2}
\frac{6\al}{\pi f(z)} F(z)
\left(
|D(m_b)|^2+A
\right)
\,,
\eeq
where $z\ev m^2_{c,pole}/m^2_{b,pole}$ and
\bea
f(z) &=&1-8z^2+8 z^3-z^4-12 z^2 \ln z
,\\ \no
F(z) &=& \frac{1}{\kappa(z)}\left(
1-\frac{8}{3}\frac{\alpha_s(m_b)}{\pi}
\right)
\,.
\eea
The CKM factor is denoted by
$\lm_i \ev V^*_{ib} V_{is} (i=u,c,t)$.
The bremsstrahlung corrections and the necessary virtual corrections,
in order to cancel the infrared divergence,
are included in the term $A$  \cite{Chetyrkin:1996vx}.

\subsection{Effects of the RS-bulk SM}

In the RS-bulk SM
the $b \to s \gamma$ decay receives
new contribution
from the KK modes of the bulk $W$ gauge boson and up-type quarks.
As discussed before, the new contribution to FCNC
occurs since
the degeneracy in the KK modes of up-type quarks
is broken due to the top quark Yukawa coupling.
We also point out that
the effects of the RS-bulk SM on FCNC
have complicated and distinct features
compared to those of other extra dimensional models.
For instance, in the models with universal flat extra dimensions
accessible to all the SM fields,
the conservation of extra dimensional momentum leads to
the so-called KK number conservation  \cite{Uni-ph},
which says that no single KK excited mode can be produced.
This reduces the computation of its contribution
to FCNC  \cite{Agashe:2001xt}.
In the RS-bulk SM, however,
non-trivial geometry does not respect this
KK number conservation with which, e.g.,
$C^{ff'W}_{00n}$ would vanish for $n\neq 0$.
Furthermore,
in universal flat extra dimensions
the bulk wave functions of bulk fermions are
the same as those of bulk gauge bosons.
The orthonormal conditions for the bulk wave functions
simplify three-point couplings so that
$C_{0nm}^{ff'W}( \textsc{flat}) = \dt_{nm}$.
In the RS model,
the bulk fermions have different bulk wave functions,
sensitively dependent on the bulk fermion mass $m_\psi$.
Non-trivial three-point couplings become involved.

Note that all the external particles are SM particles,
i.e., zero modes of the bulk fields.
Since  $\chi^{(0)}_A$,
the bulk wave function of the photon zero mode,
is constant,
the orthonormal conditions in Eqs.\,(\ref{otho-A}) and
(\ref{cond}) imply
\beq
C^{WWA}_{nm0} = \dt_{nm},\quad
C^{f\bar{f}A}_{nm0} = \dt_{nm}.
\eeq
Thus the contribution via the $n$-th up-type quark and the $m$-th $W$ boson
is the same as the SM result except for the internal mass
and the additional three-point coupling.

With a given $W^{(l)}$, the contributions of all the KK modes of
massless up and charm quarks to $(\lm_u+\lm_c) D$ in
Eq.\,(\ref{R}) are
\beq \label{u-c}
(\lm_u+\lm_c)
\left( \frac{m_W}{M_W^{(l)}} \right)^2
\sum_{\underline{i}=0}^{2n_\infty} \left( \sum_m
\Cu_{0ml} {\mathcal N}^{(0)}_{(\underline{i},\,m)} \right)^2
D(x^u_{(\underline{i},l)}) =-\lm_t
\left( \frac{m_W}{M_W^{(l)}} \right)^2
\sum_{j=1}^{n_\infty} \left(
\Cu_{0jl} \right)^2 D(x^u_{(\underline{i},l)}),
\eeq
where $
x_{(\underline{i},\, l)}^q \ev \left(
{M_q^{(\underline{i})}}/{M_W^{(l)}} \right)^2. $ For the equality
in Eq.\,(\ref{u-c}), we have employed the unitarity condition of
the CKM matrix $(\lm_u+\lm_c+\lm_t=0)$ and the exact expressions
for ${\mathcal N}^{(0)}$ in Eq.\,(\ref{N0}). In summary, the
contribution  of all the KK modes of bulk gauge bosons and up-type
quarks is taken into account by the following replacement:
\bea
\label{D}
D(x^t_{(0,0)})
\Rightarrow
D^{RS}&\equiv&
\left[
\sum_{l=0}^{n_\infty} \sum_{\underline{j}=0}^{2n_\infty}
\left( \frac{m_W}{M_W^{(l)}} \right)^2
\left( \sum_{m=0}^{n_\infty} \Cu_{0ml}{\mathcal N}_{(\underline{j},m)}
\right)^2 D(x^t_{(\underline{j},l)}) \right.
\\ \no & & ~~~~~\left.
-\sum_{j,\,l=1}^{n_\infty}
\left( \frac{m_W}{M_W^{(l)}} \right)^2
\left(
\Cu_{0jl}
\right)^2 D(x^{(0)}_{(j,l)})
\right]
\,.
\eea

In the limit where the elements of ${\mathcal M}_Y$
are much smaller than those of ${\mathcal M}_{KK}$ ($\nu \gsim -0.3$),
the substitution of Eqs.\,(\ref{mass}) and (\ref{N}) into(\ref{D})
yields, to leading order in $\dt$,
\bea
D(x^t_{(0,0)})
&\Rightarrow& \\ \no
D^{RS} &\simeq&
D(x^t_{(0,0)})
\\ \no
&& +\sum_{l=1}^{n_\infty}
\sum_{j=1}^{n_\infty}
\left( \frac{m_W}{M_W^{(l)}} \right)^2
\Big[
                    \left\{
                    \sum_{m=1}^{n_\infty} \Cu_{0ml}
                    \left(
                    \frac{\dt_{jm}}{\sqrt{2}}+\frac{n_{(j,\,m)}}{\sqrt{2}}\dt
                    \right)
                    \right\}^2
                    D\left(
                    x^{(0)}_{(j,l)}+{\hat{\chi}^2_j}\,x^{(0)}_{(j,l)}\dt/{x_f^{(j)}}
                    \right)
\\ \no
&& \qquad          + \left\{
                    \sum_{m=1}^{n_\infty} \Cu_{0ml}
                    \left(
                    -\frac{\dt_{jm}}{\sqrt{2}}+\frac{n_{(n_\infty+j,\,m)}}{\sqrt{2}}\dt
                    \right)
                    \right\}^2
                    D\left(
                    x^{(0)}_{u(j,l)}-{\hat{\chi}^2_j}\,x^{(0)}_{u(j,l)}\dt/{x_f^{(j)}}
                    \right)
\\ \no
&& \qquad         -(\Cu_{0jl})^2 D\left(
                    x^{(0)}_{(j,l)}
                    \right)
             \Big]
\\ \no &=&
D(x^t_{(0,0)}) +{\mathcal O}(\dt^2)
\,,
\eea
where we have used the approximation of
\beq
x^t_{(j,\,l)} \simeq x^{(0)}_{(j,\,l)}
\left(1 + {\hat{\chi}^2_j}/{x_f^{(j)}}\right),\quad
x^t_{(n_\infty+j,\,l)} \simeq x^{(0)}_{(j,\,l)}
\left(1 - {\hat{\chi}^2_j}/{x_f^{(j)}}\right).
\eeq
We conclude that in the limit where the Yukawa masses are smaller
than the KK masses,
the effect of the RS-bulk SM on the $Br(B\to X_s\gamma)$ vanishes to leading order.

\subsection{Numerical Results}
\label{num}

Since the new parameter $m_\psi'$, introduced
to satisfy the $\Delta\rho$ constraint,
does not affect the $b\to s\gamma$ decay,
three parameters of $m_{t,0}$, $k_{EW}$ and $\nu$
determine the contributions to $b\to s\gamma$ completely.
First let us check whether the contributions
to  $b\to s \gamma$
converge as we add more and more KK modes of $W$ boson and up-type quarks.
It is useful to introduce the cut-off on the number of the KK states,
denoted by $n_C$, and check the sensitivity of
$D^{RS}$ in Eq.\,(\ref{D}) to  $n_C$.
 $D(m_b)$ is obtained by matching Wilson coefficients
at electroweak scale and the subsequent RGE running.
For the SM contribution the matching is performed
at the next-to-leading order (NLO)
while for the RS-bulk SM effects
it is done at leading order.
Then
$D^{RS}$ gets the effects from  KK mode through
$C^{RS}_7$.
With the formula of Eq.\,(\ref{D}),
we numerically compute $C^{RS}_7$
as a function of $n_C$.
In principle we should diagonalize an infinite dimensional
matrix ${\mathcal M}_{t}$
to obtain its mass eigenstates and mixing matrix.
In reality of numerical calculation, however, this
is not possible.
Instead we take
a large finite number $n_\infty$, and truncate
the contributions of KK modes at $n_C(< n_\infty)$.

\begin{figure}
\includegraphics[height=10cm,angle=-90]{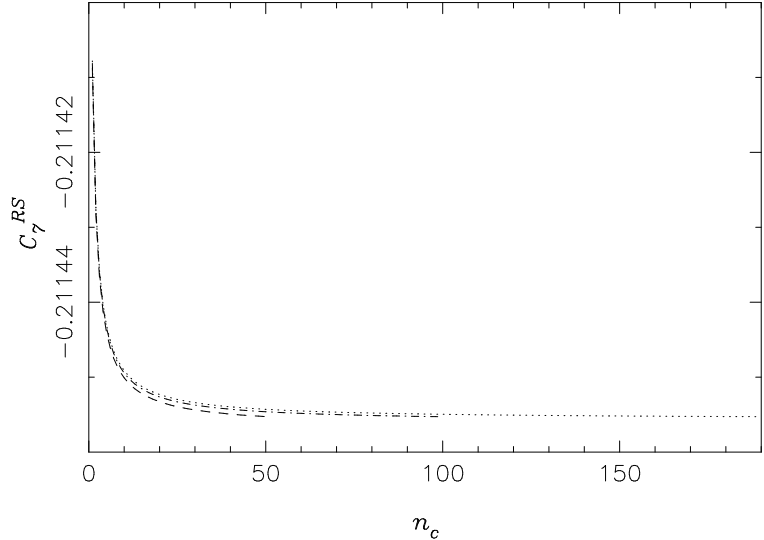}
 \caption{\label{c7}
$C^{RS}_7$ as a function of $n_c$, with $k_{EW}=10$ TeV and
$m_{t,0}=200$ GeV. The dashed line is for $n_{\infty}=50$, the
dashed-dotted line for $n_{\infty}=100$, and the dotted line for
$n_{\infty}=200$. }
\end{figure}
Figure~\ref{c7} shows $C^{RS}_7$ with the RS-bulk SM effects as a
function of $n_C$ for $k r_c=11.5, ~ \nu=-0.3$, $k_{EW}=10$ TeV,
and $m_{t,0}=200$ GeV. The dashed line is for $n_{\infty}=50$, the
dashed-dotted line for $n_{\infty}=100$, and the dotted line for
$n_{\infty}=200$. Numerical diagonalization of a large dimensional
matrix is performed by a \textsf{FORTRAN} package \textsf{LAPACK}
\cite{LAPACK}. It is clear that the $C^{RS}_7$ is well behaved
even with extremely many KK modes of $W$ boson and up-type quarks.
We note that this converging behavior is mainly due to the
suppression from the $W$ boson KK masses. Without the
factor $\left( {m_W}/{M_W^{(l)}} \right)^2$ in Eqs.~(\ref{D}), for
example, $C_7^{RS}$ at $n=n_C=50$ would not converge but keep
decreasing with the value of order -0.5. In fact the RS-bulk SM
effects on the precision electroweak observables
\cite{Davoudiasl:2000wi}, and on the anomalous magnetic moment of
the muon \cite{Davoudiasl:2000my} have been also shown finite. In
what follows we employ $n_\infty=$ 100 and $n_C=$ 50.

\begin{figure}
\includegraphics[height=10cm,angle=-90]{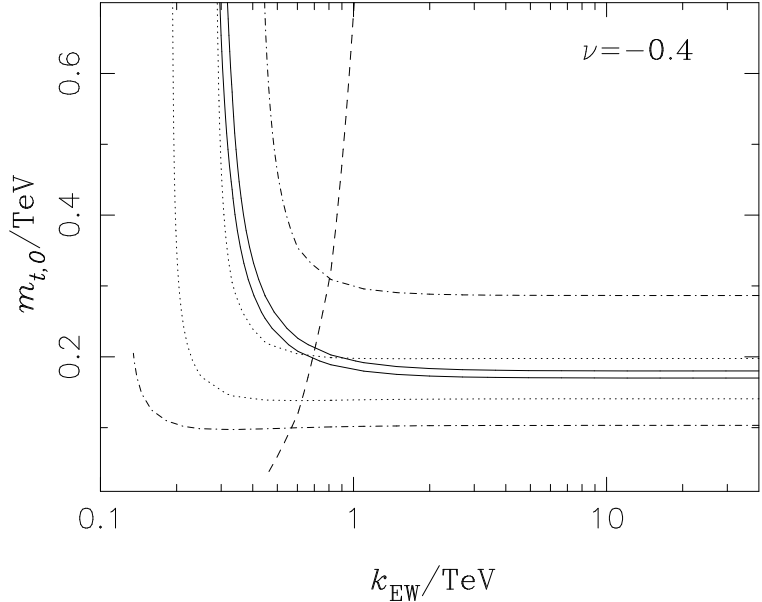}
\caption{\label{bsr4} The $(k_{EW},m_{t,0})$ parameter for  $\nu=-0.4$.
We take $k r_c=11.5$,
$n_{\infty}=100$ and $n_{c}=50$. The solid lines are from the top
mass constraints. The dashed line corresponds to $M^{\rm top}_1 \simeq 1$ TeV,
and the dash-dotted lines from the measurement of $b\rightarrow s
\gamma$ at 95\%CL.  }
\end{figure}
In Fig.\,\ref{bsr4},
we present phenomenological constraints on  $(k_{EW},m_{t,0})$
plane for $\nu=-0.4$ as suggested by the previous $\Delta\rho$ constraint.
The solid lines come from the observed top quark mass of
$175\pm 5$ GeV.
For large $k_{EW}$ of a few TeV,  $m_{t,0}$ itself is
the physical mass of the top quark,
while low $k_{EW}$ of the order 100 GeV allows considerably high
$m_0$. We notice that
the observed top quark mass alone can
put a significant lower bound on  $k_{EW}$ such as
$k_{EW} \gsim 310$~GeV for $\nu=-0.4$.
Taking a step further, let us investigate the implication of the first excited KK
mode of top quark, albeit unobserved yet.
As an example, we present the condition, denoted by the dashed lines,
that the first excited KK mode of top quark
has mass about 1~TeV, which
removes most of the parameter space
with low $k_{EW}$ and high $m_0$.
The current data
on Br($B\rightarrow X_s+\gamma)$  \cite{Buras:1998ra}
are also presented with
the doted lines at one sigma level,
and the dot-dashed lines at two sigma level.
It shows that the modified RS-bulk SM
can also accommodate the $b \to s \gamma$ constraint.
At one sigma level,
the $b \to s \gamma$ decay, with the constraint from the observed top quark mass,
can impose strong lower bound on $k_{EW}$:
$k_{EW} \gsim 700$~GeV for $\nu=-0.4$;
these correspond to the $M^{\rm top}_1 \simeq 1$ TeV.
This bound is to be compared with the constraints
from the Drell-Yan and dijet production at the Tevatron Run I,
with the oblique parameter constraints, which put the lower bound
on $m^{(1)}_{graviton} \gsim 750$ GeV \cite{Hewett:2002fe}.
Since $m^{(1)}_{graviton}\simeq 3.83 \, k_{EW}$,
this Tevetron bound is rather weak as $k_{EW} \gsim 196$ GeV.

Some comments on the $D$-meson mixing is in order here. Since our
assignment of different bulk fermion mass for the SU(2)-singlet
$b$-quark yields different KK mass spectrum of the $b$-quark from
those of the strange- and down-quark, we may encounter sizable
FCNC in the up-quark sector. Further investigation is to be done.
However, considering the current low accuracy on the $D$-meson
mixing measurement \cite{CLEO-D}, and the role of $W$ boson KK
mass spectrum in suppressing and thus stabilizing the new
contribution to $b \to s\gamma$ due to the factor $\left(
{m_W}/{M_W^{(l)}} \right)^2$, we expect that the new effects on
the $D$-meson mixing is also likely to be acceptable.

\section{Conclusions}
\label{conclusion}

We have studied the quark Kaluza-Klein mode mixing
in the Randall-Sundrum scenario where all the SM fields
except for the Higgs field are placed in the bulk.
This KK mode mixing occurs due to the Yukawa masses
of the bulk fermion with the Higgs field confined to our brane.
We have reviewed in detail the KK reduction of the bulk Dirac fermion
field in the RS background.
In order to obtain non-trivial solution of the bulk wave functions,
a Dirac fermion with a definite hypercharge should possess
both chiral states.
That is, there must be additional right-handed SU(2)--doublet
and left-handed SU(2)--singlet fermions which are to be
assigned $Z_2$-odd symmetry to avoid their zero modes
on our brane.
It is explicitly shown that
the KK mass terms are between the left- and right-handed chiral states
of a specific representation
while Yukawa couplings relate the SU(2)--doublet and SU(2)--singlet.
For the top quark of non-negligible Yukawa mass,
this mismatch between KK mass matrix and Yukawa mass matrix generates
the mixing among the KK modes of the top quark.

Immediate problems of this top quark KK mode mixing
are their dangerous effect on the $\rho$ parameter and the FCNC.
We have shown that in the minimal model with a common bulk fermion mass,
quite large mass shifts between the top and bottom quark KK modes are generated,
resulting in disastrous contribution to the $\rho$ parameter.
The minimal model is to be extended.
We relax the universal bulk fermion mass,
and let the SU(2)--singlet bottom quark field have
a different bulk fermion mass $m_\psi'$.
It is shown that for example if $m_\psi'/k \simeq -0.6$
and $m_\psi/k \simeq -0.4$,
the degeneracy of the top and bottom quark KK mode is good enough
to suppress the new contribution to $\Delta \rho$.
Moreover, the higher the KK mode is,
the better the degeneracy becomes;
the contribution of higher KK modes becomes less important.
Explicit calculation of the mass eigenvalues and mixing matrix
of top quark KK modes shows some generic features.
For $\nu \gsim -0.3$ the mixing is small such that
the diagonalization can be made perturbatively;
their contribution to the rare decay $b \to s\gamma$
vanishes to leading order.
For $\nu \lsim -0.4$ the mixing is sizable:
The diagonalization was performed numerically.
We have demonstrated that even in this case
the new effects on the $b \to s\gamma$
can be computed with high reliability.
The current measurement of Br($B\rightarrow X_s+\gamma)$
is well accommodated in the modified RS-bulk model.
If the future experiment probes the Br($B\rightarrow X_s+\gamma)$
at the current one sigma level,
this rare decay mode,
with the observed top quark mass, can put
indirect and meaningful bounds on $k_{EW}$ such as
$k_{EW} \gsim 3$~TeV for $\nu=-0.4$.

\acknowledgments
We thank G. Cvetic, H.D.~Kim and Y.Y. Keum for valuable discussions.
The work of C.S.K. was supported
in part by  CHEP-SRC Program, Grant No. 20015-111-02-2
and Grant No. R02-2002-000-00168-0 from BRP of the KOSEF,
in part by BK21 Program and Grant No. 2001-042-D00022 of the KRF.
The work of J.D.K. was supported by Grant No.
2001-042-D00022 of the KRF.
The work of J.S. was supported by Grant No. R02-2002-000-00168-0
from BRP of the KOSEF.

\def\MPL #1 #2 #3 {Mod. Phys. Lett. {\bf#1},\ #2 (#3)}
\def\NPB #1 #2 #3 {Nucl. Phys. {\bf#1},\ #2 (#3)}
\def\PLB #1 #2 #3 {Phys. Lett. {\bf#1},\ #2 (#3)}
\def\PR #1 #2 #3 {Phys. Rep. {\bf#1},\ #2 (#3)}
\def\PRD #1 #2 #3 {Phys. Rev. {\bf#1},\ #2 (#3)}
\def\PRL #1 #2 #3 {Phys. Rev. Lett. {\bf#1},\ #2 (#3)}
\def\RMP #1 #2 #3 {Rev. Mod. Phys. {\bf#1},\ #2 (#3)}
\def\NIM #1 #2 #3 {Nucl. Inst. Meth. {\bf#1},\ #2 (#3)}
\def\ZPC #1 #2 #3 {Z. Phys. {\bf#1},\ #2 (#3)}
\def\EJPC #1 #2 #3 {E. Phys. J. {\bf#1},\ #2 (#3)}
\def\IJMP #1 #2 #3 {Int. J. Mod. Phys. {\bf#1},\ #2 (#3)}
\def\JHEP #1 #2 #3 {J. High En. Phys. {\bf#1},\ #2 (#3)}
%



\begin{thebibliography}{99}

\bibitem{Antoniadis:1998ig}
N. Arkani-Hamed, S. Dimopoulos, and G. Dvali,
\PLB B429 263 1998 ;
\PRD D59 086004 1999 ;
I. Antoniadis, N. Arkani-Hamed, S. Dimopoulos, and
G. Dvali,  \PLB B436 257 1998 .

\bibitem{Randall:1999ee}
L.~Randall and R.~Sundrum,
Phys.\ Rev.\ Lett.\  {\bf 83}, 3370 (1999).

\bibitem{Dienes:1998vh}
K.~R.~Dienes, E.~Dudas and T.~Gherghetta,
Phys.\ Lett.\ B {\bf 436}, 55 (1998);
K.~R.~Dienes, E.~Dudas and T.~Gherghetta,
Nucl.\ Phys.\ B {\bf 537}, 47 (1999).

\bibitem{Antoniadis:1990ew}
I.~Antoniadis,
Phys.\ Lett.\ B {\bf 246}, 377 (1990);
T.~Han, D.~Marfatia and R.~J.~Zhang,
Phys.\ Rev.\ D {\bf 61}, 013007 (2000);
D.~E.~Kaplan, G.~D.~Kribs and M.~Schmaltz,
Phys.\ Rev.\ D {\bf 62}, 035010 (2000).

\bibitem{Arkani-Hamed:1999dc}
N.~Arkani-Hamed and M.~Schmaltz,
Phys.\ Rev.\ D {\bf 61}, 033005 (2000).

\bibitem{Arkani-Hamed:2000hv}
N.~Arkani-Hamed, H.~C.~Cheng, B.~A.~Dobrescu and L.~J.~Hall,
Phys.\ Rev.\ D {\bf 62}, 096006 (2000).

\bibitem{ADD-ph}
G.~F. Giudice, R. Rattazzi and J.~D. Wells,
\NPB B544 3 1999 ;
T.~Han, J.~D.~Lykken and R.~Zhang,
Phys. Rev. {\bf D59}, 105006 (1999);
E.~A.~Mirabelli, M.~Perelstein and M.~E.~Peskin,
Phys. Rev. Lett. {\bf 82}, 2236 (1999);
J.~L.~Hewett, Phys. Rev. Lett. {\bf 82}, 4765 (1999);
K.~Y.~Lee, H.~S.~Song and J.~Song,
Phys.\ Lett.\ B {\bf 464}, 82 (1999);
K.~Y.~Lee, H.~S.~Song, J.~Song and C.~Yu,
Phys.\ Rev.\ D {\bf 60}, 093002 (1999);
K.~Y.~Lee, S.~C.~Park, H.~S.~Song, J.~Song and C.~Yu,
Phys.\ Rev.\ D {\bf 61}, 074005 (2000);
C.~S.~Kim, K.~Y.~Lee and J.~Song,
Phys.\ Rev.\ D {\bf 64}, 015009 (2001).

\bibitem{Uni-ph}
T.~Appelquist, H.~C.~Cheng and B.~A.~Dobrescu,
Phys.\ Rev.\ D {\bf 64}, 035002 (2001).
M.~Masip and A.~Pomarol,
Phys.\ Rev.\ D {\bf 60}, 096005 (1999);
T.~G.~Rizzo and J.~D.~Wells,
Phys.\ Rev.\ D {\bf 61}, 016007 (2000).

\bibitem{Agashe:2001xt}
K.~Agashe, N.~G.~Deshpande and G.~H.~Wu,
Phys.\ Lett.\ B {\bf 514}, 309 (2001).

\bibitem{Davoudiasl:1999tf}
H.~Davoudiasl, J.~L.~Hewett and T.~G.~Rizzo,
Phys.\ Lett.\ B {\bf 473}, 43 (2000).

\bibitem{Davoudiasl:2000wi}
H.~Davoudiasl, J.~L.~Hewett and T.~G.~Rizzo,
Phys.\ Rev.\ D {\bf 63}, 075004 (2001).

\bibitem{Gherghetta:2000qt}
T.~Gherghetta and A.~Pomarol,
Nucl.\ Phys.\ B {\bf 586}, 141 (2000).

\bibitem{GIM}
S.~L. Glashow, J. Iliopoulos, and L. Maiani,
Phys.\ ReV.\ D {\bf 2}, 1285 (1970).

\bibitem{top-mixing}
F.~del Aguila, M.~P\'{e}rez-Victoria and J.~Santiago,
Phys.~Lett.~B{\bf 492}, 98 (2000);
D.~E.~Kaplan and T.~M.~Tait,
JHEP {\bf 0111}, 051 (2001).

\bibitem{Hewett:2002fe}
J.~L.~Hewett, F.~J.~Petriello and T.~G.~Rizzo,
arXiv:hep-ph/0203091.

\bibitem{Pomarol:1999ad}
A.~Pomarol,
Phys.\ Lett.\ B {\bf 486}, 153 (2000).

\bibitem{Delgado:1998qr}
A.~Delgado, A.~Pomarol and M.~Quiros,
Phys.\ Rev.\ D {\bf 60}, 095008 (1999).

\bibitem{Grossman:1999ra}
Y.~Grossman and M.~Neubert,
Phys.\ Lett.\ B {\bf 474}, 361 (2000).

\bibitem{Chang:1999nh}
S.~Chang, J.~Hisano, H.~Nakano, N.~Okada and M.~Yamaguchi,
Phys.\ Rev.\ D {\bf 62}, 084025 (2000).

\bibitem{Davoudiasl:2000my}
H.~Davoudiasl, J.~L.~Hewett and T.~G.~Rizzo,
Phys.\ Lett.\ B {\bf 493}, 135 (2000).

\bibitem{Greub:1997hf}
C.~Greub, T.~Hurth and D.~Wyler,
Phys.\ Lett.\ B {\bf 380}, 385 (1996);
C.~Greub and T.~Hurth,
Phys.\ Rev.\ D {\bf 56}, 2934 (1997);
H.~Baer and M.~Brhlik,
Phys.\ Rev.\ D {\bf 55}, 3201 (1997).

\bibitem{PDG}
D.~E.~Groom {\it et al.}  [Particle Data Group Collaboration],
Eur.\ Phys.\ J.\ C {\bf 15} (2000) 1.

\bibitem{Buras:1998ra}
A.~J.~Buras,
arXiv:hep-ph/9806471.

\bibitem{Chetyrkin:1996vx}
K.~Chetyrkin, M.~Misiak and M.~Munz,
Phys.\ Lett.\ B {\bf 400}, 206 (1997)
[Erratum-ibid.\ B {\bf 425}, 414 (1997)].

\bibitem{Davoudiasl:1999jd}
H.~Davoudiasl, J.~L.~Hewett and T.~G.~Rizzo,
Phys.\ Rev.\ Lett.\  {\bf 84}, 2080 (2000).

\bibitem{LAPACK}
\textsf{http://www.netlib.org/lapack/}

\bibitem{CLEO-D}
CLEO Collaboration (Alex B. Smith for the collaboration):
arXiv:hep-ex/0206001.


\end{thebibliography}
\end{document}